\documentclass[3p]{elsarticle}
\usepackage{amssymb}
\usepackage[utf8]{inputenc}
\usepackage{graphicx}
\usepackage{bm,color,subfigure,amsmath,hyperref}
\begin{document}

\title{Perspectives on spin hydrodynamics in ferromagnetic materials}
% 6 pages with references!

\author[appm]{Ezio~Iacocca}
\ead{ezio.iacocca@colorado.edu}

%\author[nist]{T. J.~Silva}

\author[appm]{Mark~A.~Hoefer}

\address[appm]{Department of Applied Mathematics, University of Colorado, Boulder, Colorado 80309-0526, USA}
%\address[nist]{National Institute of Standards and Technology, Boulder, Colorado 80305-3328, USA}

\begin{abstract}
  The field of spin hydrodynamics aims to describe magnetization
  dynamics from a fluid perspective. For ferromagnetic materials,
  there is an exact mapping between the Landau-Lifshitz equation and a
  set of dispersive hydrodynamic equations. This analogy provides
  ample opportunities to explore novel magnetization dynamics and
  magnetization states that can lead to applications relying entirely
  upon magnetic materials, for example, long-distance transport of
  information. This article provides an overview of the theoretical
  foundations of spin hydrodynamics and their physical interpretation
  in the context of spin transport. We discuss other proposed
  applications for spin hydrodynamics as well as our view on
  challenges and future research directions.
\end{abstract}

\maketitle

% FIG 1: SDW and hydrodynamic representation. Adapt from ppt.
% FIG 2: (a) n vs u stability (b) v-av (from prb & kim)
% FIG 3: Transport (take from published) (a) General setup (b) below a DEF, CS-DEF, and sym-broken (c) frequencies

\section{Introduction}

%Don't write about spin currents extensively. Probably the first paragraph can be deleted.

% Introductory sentence on spins and technology associated with it
%An important milestone at the turn of the millennium was the discovery of spin-transfer torque (STT)~\cite{Slonczewski1996,Berger1996,Tsoi1998}. STT describes the mechanism by which spin can be transferred onto the magnetization at a magnetic interface via angular momentum conservation~\cite{Ralph2008}. While STT can be described from a quantum mechanical perspective~\cite{Slonczewski1999}, sizable effects can be only observed when a large amount of spins with the same orientation impinge on the magnetic interface. This condition was first realized via spin-polarized currents generated by the spin-filter effect~\cite{Toi1998?}, where the spin of electrons conducting through a magnetic material aligns with the magnetization in majority and minority channels~\cite{Stohr2006}. STT has been successful in a variety of settings, leading to the development of magnetic memory elements~\cite{?}, oscillators~\cite{Chen2016b}, and new measurement techniques~\cite{?}.

% Introduction to spin currents
A new paradigm for magnetic-based technologies is to expand the
functionality and energy efficiency of spintronic
devices~\cite{Roadmap2017} by utilizing spin
currents~\cite{Hoffmann2007,Jungwirth2012}. Spin currents describe the
transport of angular momentum by particles or
quasi-particles. % and can be represented by a continuity equation of the spin density
%\begin{equation}
%\label{eq:spincurrent}
%  \partial_t\mathbf{s} + \nabla\cdot\bar{\mathbf{Q}}_s = \frac{1}{T},
%\end{equation}
%where $\mathbf{s}=(s_x,s_y,s_z)$ is the spin density vector and $\bar{\mathbf{Q}}_s$ is the spin current tensor, indicating three-dimensional transport for each spin component. Spin currents are not conserved, so that the continuity equation is subject to a spin relaxation term $1/T$~\cite{Zhang2005}.
% Spin currents in metals
The quintessential spin-carrying particle is the electron. Spin currents carried by electrons can be generated %in metals exhibiting Rashba or Dresselhaus spin-orbit coupling that locks spin {and momentum} at the Fermi energy~\cite{Manchon2015}; or metals containing spin-orbit scatterers, i.e, heavy metals or heavy metal defects, resulting in the spin Hall effect~\cite{Hoffmann2013}
in nonmagnetic~\cite{Manchon2015,Hoffmann2013} and magnetic~\cite{Humphries2017,Kimata2019} metals, but are necessarily accompanied by %. In other words, spin currents carried by electrons require 
a net charge current. This charge current incurs energy dissipation via Joule heating that limits the energy efficiency of charge-to-spin interconversion. Despite this {shortcoming}, spin currents hold promise for {current-controlled magnetization dynamics and their applications~\cite{Jungwirth2012}}

% Spin currents via magnons
Spin waves are fundamental magnetic excitations that also carry spin current as quantum-mechanical quasi-particles known as magnons~\cite{White2007}. %Although magnons obey Bose-Einstein statistics~\cite{Demokritov2006}
In the context of spin transport, the spin transfer torque~\cite{Slonczewski1996,berger1996} and the spin pumping~\cite{Tserkovnyak2002b,Brataas2012} effects provide an interconversion mechanism between angular momentum and charge currents at a magnet / metal interface. %Therefore, magnons can be used to transport spin through magnetic materials.
A technological advantage of spin waves is that their existence is independent of the conduction properties of the material. This implies that energy dissipation associated to conduction electrons is minimized. % by utilizing insulators such as yttrium-iron-garnet~\cite{?} or metallic alloys exhibiting a semi-metallic band structure~\cite{Schoen2016}. However, energy can be redistributed via magnon scattering and ultimately dissipated via magnon-phonon interactions~\cite{Suhl1998}, limiting the spin current transport capacity of magnons.
However, spin waves are subject to scattering processes that ultimately limit their coherence~\cite{Suhl1998}.

% Long-range transport of magnons in AFMs
%In Eq.~\eqref{eq:LLG}, magnons are equivalent to spin waves defined as small-amplitude perturbations of a uniform magnetic state~\cite{White2007}. 
The amplitude of a spin wave decays exponentially with a decay length equal to $v_g/(2\omega\alpha)$~\cite{Madami2011}, where $v_g$, $\omega$, and $\alpha$ are, respectively, the spin wave group velocity, angular frequency, and the magnetic {Gilbert} damping parameter~\cite{Gilbert2004}. To maximize the decay length, low damping materials such as Permalloy (Ni$_{80}$Fe$_{20}$) and YIG %can maximize the decay length on the order of micrometers~\cite{Madami2011,Cornelissen2015} and
have been regularly used for research on all-magnetic logic and computation~\cite{Chumak2015}. Decay lengths on the order of micrometers have been obtained in these materials~\cite{Madami2011,Cornelissen2015,Liu2018}. %More recently, similar large decay lengths have been observed for short-wavelength spin waves~\cite{Liu2018}.
Recently, long-distance spin transport in amorphous Yttrium-Iron ferrite~\cite{Wesenberg2017} and antiferromagnetic haematite~\cite{Lebrun2018} was measured experimentally.
%
% Spin transport through noncollinear states
%Despite the notable advances in achieving long-distance spin transport via spin waves, the 
 {However, the} finite Gilbert damping parameter and the concomitant exponential decay of spin waves remains a presently insurmountable limitation for magnon-based technologies.% {[I think it is reasonable that damping cannot be removed and spin waves will decay]}.

 To beat exponential decay, other forms of spin transport must be
 explored. Spin hydrodynamics offers such an alternative via the
 stabilization of noncollinear, large-amplitude magnetization states
 in magnetic materials with dominant easy-plane anisotropy. Because of
 its potential impact on technologies that rely on the control of
 spin, the field of spin hydrodynamics has experienced a rapid growth
 over the last five years, including recent experimental
 evidence~\cite{Stepanov2018,Yuan2018} for hydrodynamic-like spin
 transport.

% Concluding paragraph: in this review... talk about the math, the fluid description and the relevance of nonlinear effects
In light of multiple recent theoretical developments, we provide in this article an overview of the theoretical studies on spin hydrodynamics in ferromagnetic materials. We review the dispersive hydrodynamic formulation of magnetization dynamics and its physical interpretation. We also discuss the theoretical predictions pertaining to the properties of ideal and current-induced spin hydrodynamic states in the context of long-distance spin transport. {Finally, we discuss challenges and possible research directions.}

\section{Fluid interpretation of magnetization dynamics}

\subsection{Micromagnetic equations of motion}

The dynamics of the ferromagnetic space- and time-dependent magnetization vector $\mathbf{M}$ can be described by the Landau-Lifshitz (LL) equation~\cite{Landau1953}
\begin{equation}
\label{eq:LLG}
  \partial_t\mathbf{M} = -\gamma\mu_0\left[\mathbf{M}\times\mathbf{H}_\mathrm{eff} + \frac{\alpha}{M_s}\mathbf{M}\times\left(\mathbf{M}\times\mathbf{H}_\mathrm{eff}\right)\right],
\end{equation}
where $M_s=|\mathbf{M}|$ is the saturation magnetization, $\gamma$ is the gyromagnetic ratio, $\mu_0$ is the vacuum permeability, and $\alpha$ %is the phenomenological Gilbert damping parameter~\cite{Gilbert2004} that
can be {utilized} %written 
in the LL form of Eq.~\eqref{eq:LLG} when $\alpha\ll1$. $\mathbf{H}_\mathrm{eff}$ is an effective field that models the relevant physics acting on the magnetic material. For simplicity, we primarily consider thin film materials with dominant uniaxial symmetry so that the effective field includes an external field, uniaxial anisotropy, and exchange,
\begin{equation}
\label{eq:Heff_ferro}
\mathbf{H}_\mathrm{eff} =  \underbrace{\mathbf{H}_0}_\text{external field}-\underbrace{\left(1-h_k\right)(\mathbf{M}\cdot\hat{\mathbf{z}})\hat{\mathbf{z}}}_\text{uniaxial anisotropy}+\underbrace{\lambda_\mathrm{ex}^2\Delta\mathbf{M}}_\text{exchange}.
\end{equation}

Here, $h_k=H_k/M_s$ is a perpendicular anisotropy field that can arise, e.g., from surface effects in multilayers~\cite{Bruno1989}; the $(-\mathbf{M}\cdot\hat{\mathbf{z}})\hat{\mathbf{z}}$ term corresponds to the local demagnetizing field in a thin film with a uniform magnetization in the perpendicular-to-plane direction $\hat{\mathbf{z}}$; and $\lambda_\mathrm{ex}=\sqrt{2A/(\mu_0M_s^2)}$ is the exchange length with $A$ being the exchange constant.

We can identify two fundamentally distinct hydrodynamic regimes by introducing the quantity $\sigma=\mathrm{sgn}(1-h_k)$. The sign $\sigma=1$ represents materials with easy-plane anisotropy, e.g. Py, which we will show to be hydrodynamically \textit{stable}, and $\sigma=-1$ represents materials with perpendicular magnetic anisotropy, e.g., Co/Ni multilayers, which are hydrodynamically \textit{unstable}. The case $\sigma=0$, i.e., $h_k=1$, represents isotropic ferromagnets that are not considered here.

To simplify notation, it is convenient to work with the dimensionless LL equation~\cite{Iacocca2017}
\begin{subequations}
\begin{eqnarray}
\label{eq:ll}
  \partial_t\mathbf{m} &=& -\mathbf{m}\times\mathbf{h}_\mathrm{eff} - \alpha\mathbf{m}\times\left(\mathbf{m}\times\mathbf{h}_\mathrm{eff}\right),\\
\label{eq:heff}
	\mathbf{h}_\mathrm{eff} &=&  \mathbf{h}_0\hat{\mathbf{z}}-\sigma(\mathbf{m}\cdot\hat{\mathbf{z}})\hat{\mathbf{z}}+\Delta\mathbf{m},
\end{eqnarray}
\end{subequations}
where $\mathbf{m}=\mathbf{M}/M_s$, fields are scaled by $|1-h_k|M_s$, space is scaled by $\sqrt{|1-h_k|}\lambda_\mathrm{ex}^{-1}$, and time is scaled by $\gamma\mu_0M_s|1-h_k|$.

%For antiferromagnets, magnetization dynamics are typically obtained by considering two ferromagnetic sub-lattices, extending the formulation as a set of two, coupled LL equations. For a recent review see, e.g., Ref.~\cite{Baltz2018}.

\subsection{Hydrodynamic transformation}

%o	Transformation
The fluid interpretation of the LL equation, Eq.~\eqref{eq:LLG}, was first proposed by Halperin and Hohenberg~\cite{Halperin1969} in the context of spin wave dispersion. %More generally, a hydrodynamic formulation of the LL equation is achieved by utilizing a canonical transformation between the magnetization vector and a spin density and fluid velocity. 
For %the case of 
an easy-plane ferromagnet, a %convenient transformation is
canonical transformation between the magnetization vector and a spin density and fluid velocity is~\cite{papanicolaou_dynamics_1991}
\begin{subequations}
\label{eq:coord}
	\begin{eqnarray}
  \label{eq:coord_n}
	n &=& m_z,\\
	\label{eq:coord_u}
  \mathbf{u} &=& -\nabla \Phi = -\nabla \left [
    \arctan{\left(m_y/m_x\right)} \right ],
\end{eqnarray}
\end{subequations}
where the normalized magnetization is $\mathbf{m}=(m_x,m_y,m_y)$, $n$ is the longitudinal spin density, and $\mathbf{u}$ is the fluid velocity equal to the negative gradient of the azimuthal angle $\Phi$. %By use of the transformation~\eqref{eq:coord}, the LLG equation exactly maps into a system of dispersive hydrodynamic equations~\cite{Iacocca2017,Iacocca2017b} akin to the Euler equations of a compressible, isentropic fluid~\cite{?}. A similar transformation has been used for ferromagnets and antiferromagnets in the small-density limit~\cite{Sonin2010,Takei2014,Takei2014b,Sonin2017}, so that the nonlinear terms in the LLG equations are negligible.

%o	What is the physical meaning of the fluid representation
The spin density defined in Eq.~\eqref{eq:coord_n} is bounded by the magnetization vector magnitude, $|n|\leq1$. In contrast with typical fluids, the spin density is {also} signed. % and {bounded}. 
When the density is either $+1$ or $-1$, the azimuthal angle $\Phi$, hence the fluid velocity $\mathbf{u}$, is not defined. This implies that $n=\pm1$ corresponds to vacuum~\cite{Iacocca2017}. In contrast, $n=0$ corresponds to density saturation. %As we discuss below, fluid vacuum and saturation are intimately related to the spin current carried by spin hydrodynamics.

The fluid velocity given in Eq.~\eqref{eq:coord_u} is irrotational, i.e., $\nabla\times\mathbf{u}=\nabla\times(-\nabla\Phi)=0$ in non-vacuum regions where $\Phi$ is defined. A direct consequence is that vortices can only appear in a quantized manner via a density vacuum point~\cite{Sonin2010}. %This is just a restatement from a fluid perspective that magnetic vortices in easy-plane ferromagnets possess a core that points exactly normal to the plane~\cite{Coey2010}.
Theoretically, the fluid velocity is unbounded. However, because the transformation \eqref{eq:coord} is applied to the LL equation in a micromagnetic approximation, the concept of a fluid velocity is valid insofar as $|\mathbf{u}|\lesssim|1-h_k|^{-1/2}$, {corresponding to physical length scales larger than or on the order of $\lambda_\mathrm{ex}$}.

\begin{figure}[t]
\centering \includegraphics[trim={1in 0.6in 6.5in 0.25in}, clip, width=2.5in]{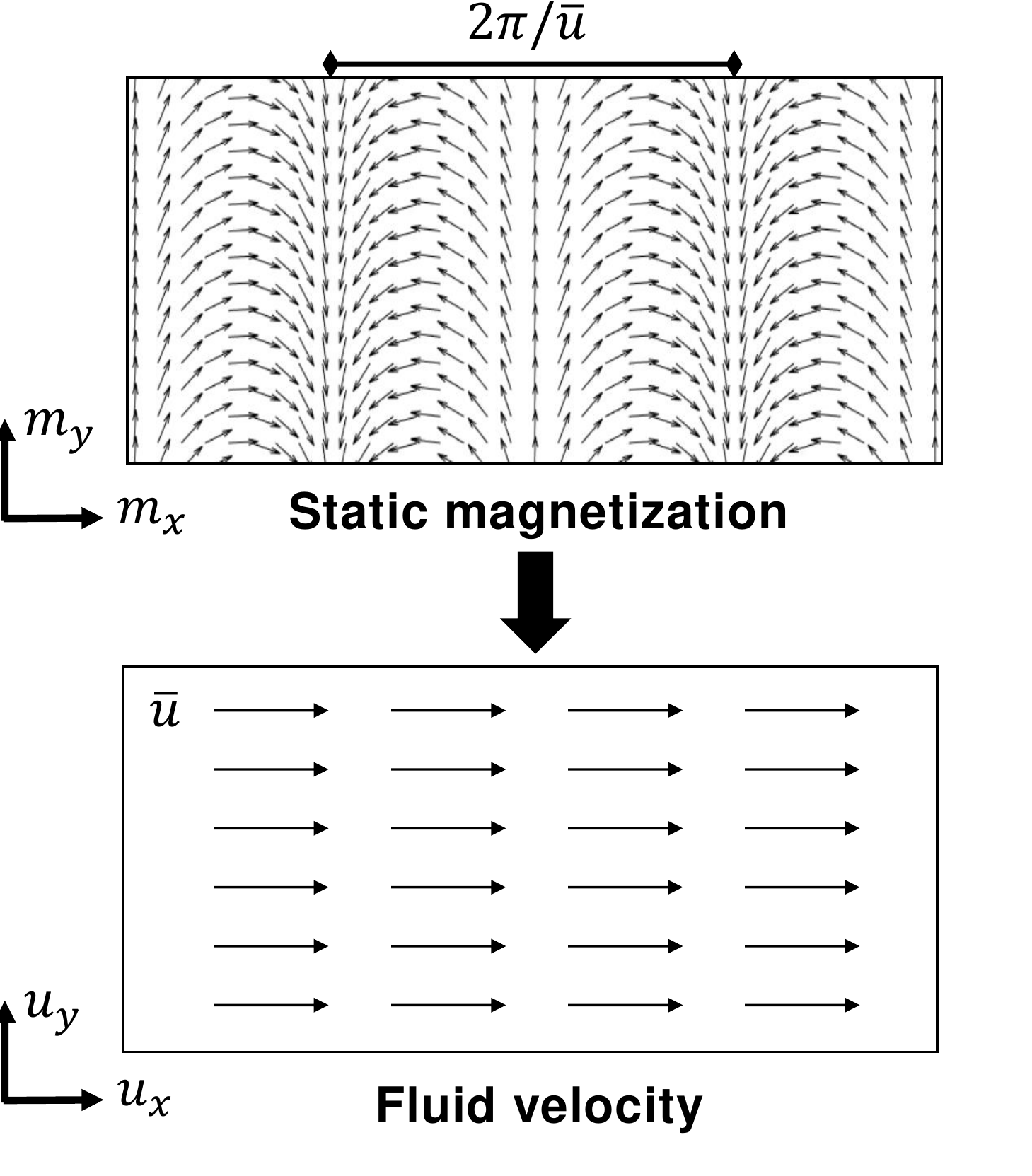}
\caption{ \label{fig1} Spin-density wave exhibiting a static, spiraling magnetization texture for the in-plane magnetization, $m_x$ and $m_y$. The equivalent fluid representation is a uniform flow $\mathbf{u}$. The fluid velocity is equivalent to the spin-density wave wavevector, as depicted in the top panel. }
\end{figure}

To better illustrate the physical meaning of the fluid velocity, it is insightful to consider the simple case of a spin supercurrent~\cite{Konig2001} where {$n=0$ and} the azimuthal angle is a linear function of space, e.g., $\Phi=-\bar{u} x$. {This describes a spiraling, static magnetization texture with a constant phase difference between neighboring magnetization vectors} similar to a spin-density wave (SDW)~\cite{Gruner1994} as depicted in Fig.~\ref{fig1}. The fluid velocity is simply equivalent to the SDW wavenumber. Note that we have not considered any time dependence. As such, the fluid velocity can be nonzero even for a static texture such as a SDW~\cite{Iacocca2017,Iacocca2017b}. {We stress that SDWs are defined for any $n=\bar{n}$ that can be stabilized by an external field $\mathbf{h}_0=-\bar{n}\hat{\mathbf{z}}$ for $|\bar{n}|<1$~\cite{Iacocca2017b}.}
%-------------------------------
%-------------------------------

\section{Noncollinear magnetization states}

%o	DH equations
\subsection{Dispersive hydrodynamic formulation}

Performing the transformation~\eqref{eq:coord} on the dimensionless LL equation~\eqref{eq:ll} and \eqref{eq:heff} yields a set of dispersive hydrodynamic (DH) equations~\cite{Iacocca2017,Iacocca2017b}
\begin{subequations}
\label{eq:nudot}
\begin{eqnarray}
  \label{eq:ndot}
    \frac{\partial n}{\partial t} &=&
    \nabla\cdot\underbrace{\left[(1-n^2)\mathbf{u}\right]}_\text{spin density
      flux} + 
    \underbrace{\alpha(1-n^2)\frac{\partial \Phi}{\partial
        t}}_\text{spin relaxation} ,\\
  \label{eq:udot}
  \frac{\partial \mathbf{u}}{\partial t} &=&
  \nabla\underbrace{\left[(\sigma-|\mathbf{u}|^2)n
    \right]}_{\text{velocity flux}} 
  - \underbrace{\nabla\left[\frac{1}{\sqrt{1-n^2}}\nabla\cdot\left(\frac{\nabla n}{\sqrt{1-n^2}}\right)\right]}_\text{dispersion}\\  
  &&\underbrace{-\nabla
    h_0}_{\text{potential force}}
  +\underbrace{\alpha\nabla\left[\frac{1}{1-n^2}\nabla\cdot\left[(1-n^2) 
        \mathbf{u}\right]\right]}_\text{viscous loss}.\nonumber
\end{eqnarray}
\end{subequations}

These equations are an exact transformation of the LL equation. {The fluid velocity Eq.~\eqref{eq:udot} was derived by taking the gradient of the phase or potential equation}
\begin{eqnarray}
\label{eq:phase}
  \frac{\partial\Phi}{\partial t} &=& -(\sigma-|\mathbf{u}|^2)n+\frac{1}{\sqrt{1-n^2}}\nabla\cdot\left(\frac{\nabla n}{\sqrt{1-n^2}}\right)\nonumber\\&+&h_0-\frac{\alpha}{1-n^2}\nabla\cdot\left[(1-n^2)\mathbf{u}\right],
\end{eqnarray}
{which, when $\alpha=0$, can be interpreted as the magnetic analogue of Bernoulli's equation.}

The physical interpretation of each term {in Eqs.~\eqref{eq:ndot} and \eqref{eq:udot} is} specified. Exchange results in long-wave spin density and velocity fluxes in Eqs.~\eqref{eq:ndot} and \eqref{eq:udot}. These types of long-wave nonlinear effects are well-known in fluid dynamics to give rise to self-steepening and shock formation~\cite{Landau1987}. Exchange dispersion in Eq.~\eqref{eq:udot} becomes more pronounced for rapidly varying disturbances, e.g., when self-steepening occurs. The anisotropy factor $\sigma$ only appears in the velocity flux term in Eq.~\eqref{eq:udot}. Losses are all proportional to $\alpha$ and have intriguing fluid interpretations. For example, the spin relaxation term drives the spin density to the equilibrium configuration $n=h_0/(\sigma-\bar{u}^2)$~\cite{Iacocca2017b}. The viscous loss term is the magnetic analogue of viscous friction in a Newtonian fluid.

%Equation~\eqref{eq:ndot} is equivalent to Eq.~\eqref{eq:spincurrent} for the $\hat{\mathbf{z}}$ spin component. Therefore, t
The first term on the right hand side {in Eq.~\eqref{eq:ndot}} includes the flux of angular momentum or \textit{spin current density} in the $\hat{\mathbf{z}}$ direction,
\begin{equation}
  \label{eq:Js}
  %\bar{\mathbf{Q}}_s\cdot\hat{\mathbf{z}}=\mathbf{J}_s \propto -(1-n^2)\mathbf{u} .
	\mathbf{Q}_s = -\frac{\mu_0M_s^2\lambda_\mathrm{ex}}{\sqrt{|1-h_k|}}(1-n^2)\mathbf{u},
\end{equation}
expressed here in units of J/m$^2$.

In the context of spin hydrodynamics, Eq.~\eqref{eq:Js} embodies the physical understanding of the fluid variables discussed above. First, the spin density sets the basis direction for the component of spin transport. Second, the prefactor $1-n^2$ is consistent with the interpretation that a maximal spin current is obtained for saturation, $n=0$, while no spin current exists at vacuum, $n=\pm1$. Third, the fluid velocity is directly proportional to the spin current magnitude and describes its flow direction. The same conclusions can be obtained from the point of view of equilibrium spin currents between exchange-coupled spins in a continuum approximation~\cite{Bruno2005}.

It is important to stress that the spin current in Eq.~\eqref{eq:Js} is related to hydrodynamic quantities that evolve in time via the DH equations~\eqref{eq:ndot} and \eqref{eq:udot}. Therefore, the spin current also evolves in time. This feature was recently recognized to describe the far-from-equilibrium transfer of angular momentum in the picosecond evolution of the spatial magnetization in GdFeCo ferrimagnetic amorphous thin films subject to ultrafast optical pumping~\cite{Iacocca2019}.

%o	Limits to BEC and spin superfluid
\subsubsection{Superfluid limit}

The DH equations~\eqref{eq:ndot} and \eqref{eq:udot} can be simplified in the long wavelength, in-plane limit, when $|n|\ll1$, $|\mathbf{u}|^2\ll1$, $|\nabla n|^2\ll 1$, and $|\Delta n|\ll1$. The resulting equations are
\begin{subequations}
\label{eq:nuss}
\begin{eqnarray}
  \label{eq:nss}
    \frac{\partial n}{\partial t} &=& \nabla\cdot\mathbf{u} + \alpha\frac{\partial \Phi}{\partial
        t},\\
  \label{eq:uss}
  \frac{\partial \mathbf{u}}{\partial t} &=&
  \sigma\nabla n -\nabla h_0 + \alpha\nabla(\nabla\cdot\mathbf{u}).
\end{eqnarray}
\end{subequations}

%Despite the presence of energy dissipation via damping, these equations have been identified to support so-called \emph{spin superfluids} in easy-plane ferromagnets ($\sigma=-1$)~\cite{Sonin2010,Takei2014,Chen2014,Skarsvaag2015,Takei2015,Flebus2016,Sonin2017,Schneider2018,Sonin2019}.
In the idealized scenario of a static texture with constant fluid velocity ($\partial\Phi/\partial t=0$ and $\nabla\cdot\mathbf{u}=0$), the damping terms are eliminated from Eqs.~\eqref{eq:nss} and \eqref{eq:uss}. This scenario corresponds to the \emph{dissipationless spin transport} identified in Ref.~\cite{Konig2001}. These equations also support so-called spin superfluids when $\partial\Phi/\partial t\neq0$~\cite{Sonin2010}. The analogy between a spin superfluid and superfluid mass transport or charge supercurrent is based on topological arguments: the noncollinear state is mapped into an order parameter that exhibits a nonzero winding number and is formally identical to superfluidity. However, any magnetization dynamics will dissipate energy via damping, as is clear from the viscous terms in Eq.~\eqref{eq:uss} proportional to $\alpha$.% From this point of view, spin superfluidity is notably different from superfluidity in which the host {magnetic material damping parameter is always capable of dissipating energy [we might remove this sentence since there are many subtleties and we already wrote the main point: damping will dissipate energy]}. %Nevertheless, the nonzero winding number offers a wealth of benefits in the context of spin transport.

\subsubsection{Bose-Einstein condensate limit}

Another interesting limit is that of a perpendicularly magnetized easy-plane ferromagnet ($\sigma=1$) subject to small perturbations. In this case, exchange spin waves~\cite{Stancil2009} are excited. The DH equations can be rewritten as a function of the spin density deviation $\rho=1-n$, where $n$ is close to 1 (vacuum). The resulting equations, upon renormalization of fluid velocity, space, time, and field read~\cite{Iacocca2017b}
\begin{subequations}
\label{eq:nubec}
\begin{eqnarray}
  \label{eq:nbec}
    \frac{\partial \rho}{\partial t} + \nabla\cdot\left(\rho\mathbf{u}\right) &=& 0,\\
  \label{eq:ubec}
  \frac{\partial\mathbf{u}}{\partial t} + \left(\mathbf{u}\cdot\nabla\right)\mathbf{u}+\nabla\rho &=& \frac{1}{4}\nabla\left[\frac{\Delta\rho}{\rho}-\frac{|\nabla\rho|^2}{2\rho^2}\right]-\nabla h_0,
\end{eqnarray}
\end{subequations}
%These equations 
{and} describe a repulsive Bose-Einstein condensate (BEC) with
trapping potential $U=h_0$~\cite{Pethick2002}. Physically, this
analogy corresponds to the bosonic character of magnons~\cite{White2007}.

\subsubsection{Two-component Bose-Einstein condensate limit}

The conservative, one-dimensional reduction of Eqs.~\eqref{eq:ndot} and \eqref{eq:udot}, i.e., when $\nabla\rightarrow\partial_x$, $\mathbf{u}=u\cdot\hat{\mathbf{x}}$, and $\alpha=0$, coincides exactly with the equations describing polarization waves in one-dimensional, two-component BECs~\cite{Qu2016,Congy2016}. These equations have been recently shown to support many interesting nonlinear wave phenomena including dispersive shock waves~\cite{El2016} from a fast spatial transition~\cite{Ivanov2017,Ivanov2018}.

%o	Simplest mathematical solution: UHS and SDWs
\subsection{Uniform solutions and topology}

The simplest solution to Eqs.~\eqref{eq:ndot} and \eqref{eq:udot} are the family of static, constant density and fluid velocity states, ($\bar{n},\bar{u}$),
\begin{equation}
\label{eq:sdw}
  n = \bar{n}=\frac{h_0}{\sigma-\bar{u}^2},\quad \mathbf{u}=\bar{u}\hat{\boldsymbol{\xi}},\quad\partial\Phi/\partial t=0,
\end{equation}
where $\hat{\boldsymbol{\xi}}$ is the flow direction in the film's plane. These solutions are SDWs that exhibit dissipationless spin transport~\cite{Konig2001} as described before.%, as represented in Fig.~\ref{fig1}. Because SDWs are static, energy dissipation via damping is inoperative.

In the conservative limit, when $\alpha=0$, SDWs can be dynamic, precessional textures. These solutions have been generically termed uniform hydrodynamic states (UHSs)~\cite{Iacocca2017,Iacocca2017b} that exhibit a spatially-uniform precessional frequency% is obtained from Eq.~\eqref{eq:udot} as
\begin{equation}
\label{eq:uhs_freq}
  \partial_t\Phi=\Omega=-(\sigma-\bar{u}^2)\bar{n}+h_0
\end{equation}

Both SDWs and UHSs exhibit a single-chirality spatial rotation along the $\hat{\mathbf{z}}$ axis. This implies that both textures are topologically protected via a nonzero winding number~\cite{Iacocca2017,Sonin2010}.

\subsection{Spin wave dispersion and stability}

%o	Stability: spin wave dispersion
The stability of SDWs and UHSs ultimately concerns the manner in which perturbations to these solutions evolve. Generically, small amplitude perturbations are spin waves that carry energy away from the host SDW or UHS. If perturbations grow in time or the generation of spin waves is energetically favorable, then the host state is unstable. This latter rationale was proposed by Landau in his criterion for superfluidity~\cite{Landau1941} and has been used in the spin superfluid limit~\cite{Sonin2010,Sonin2017,Sonin2019}.

A fluid-centric interpretation %the criterion can be found %in the subsonic to supersonic flow transition. For this, the Mach numbers are calculated from the spin wave dispersion. For spin waves on a UHS background, the dispersion is~\cite{Iacocca2017,Iacocca2017b}
{of stability criteria can be found by analyzing the spin wave dispersion relation on a UHS background~\cite{Iacocca2017,Iacocca2017b}}
\begin{equation}
  \label{eq:dispersion}
  \omega_\pm(\mathbf{k}) = 2 \bar{n} \bar{u}\hat{\boldsymbol{\xi}}  \cdot \mathbf{k} \pm |\mathbf{k}|\sqrt{(1-\bar{n}^2)
    (\sigma-\bar{u}^2)+|\mathbf{k}|^2}, 
\end{equation}
where $\mathbf{k}$ is the wave vector. In the long wavelength limit, $|\mathbf{k}|\rightarrow 0${, and if $\hat{\boldsymbol{\xi}}=\mathbf{k}/|\mathbf{k}|$}, coincident phase and group velocities allow one to identify the speeds of sound
\begin{equation}
  \label{eq:speed}
  s_\pm = 2\bar{n} \bar{u} \pm \sqrt{(1 - \bar{n}^2)(\sigma
    - \bar{u}^2)} .
\end{equation}

\subsubsection{Modulationally unstable regime}

When the speeds in Eq.~\eqref{eq:speed} are \emph{complex}, small
fluctuations can grow exponentially. As we now demonstrate, this
instability is known as modulational instability
(MI)~\cite{Whitham1974,zakharov_modulation_2009}.

\begin{figure}[t]
\centering \includegraphics[trim={0in 0in 0in 0in}, clip, width=3in]{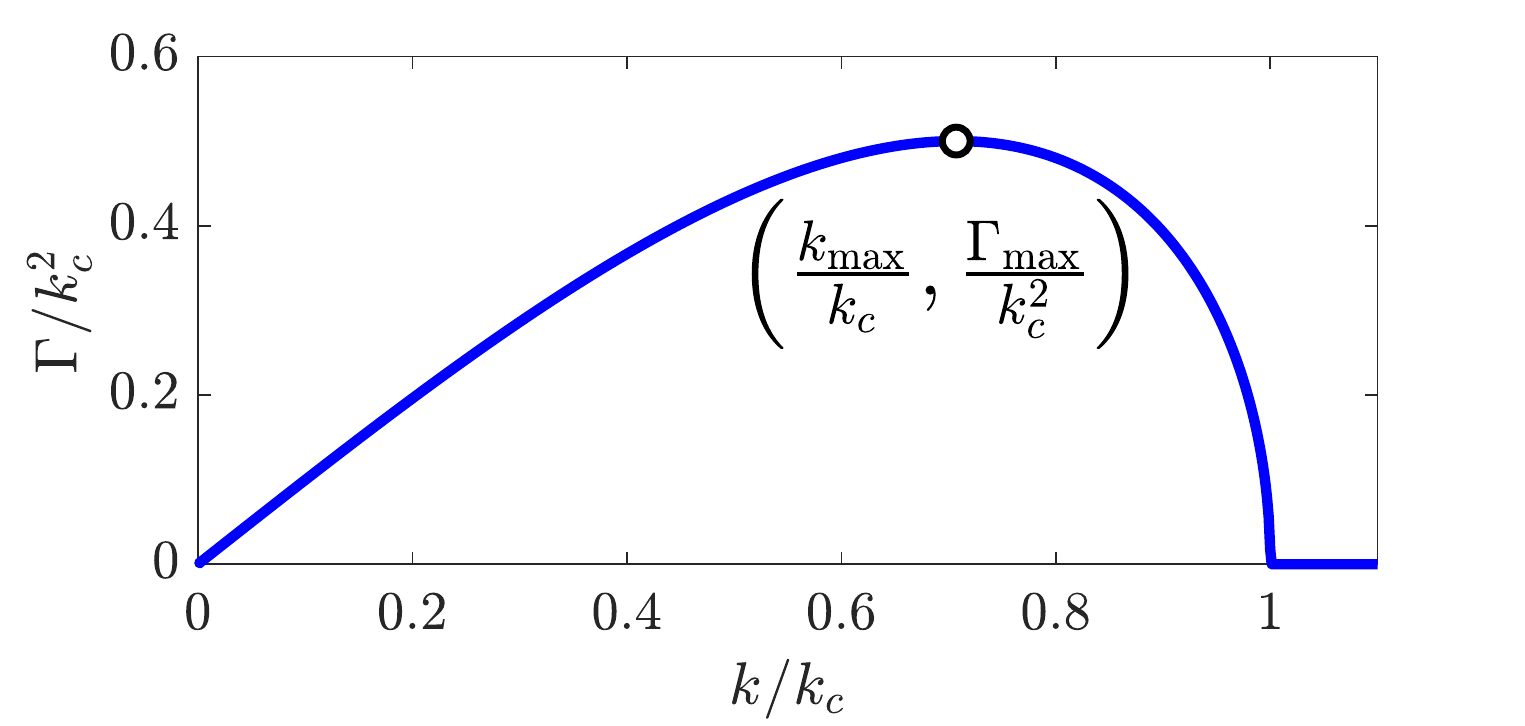}
\caption{ \label{fig_extra} Growth rate of modulationally unstable perturbations. The instability is limited to long waves, with wavevectors below $k_c$, and exhibits a maximum at the coordinates indicated in Eq.~\eqref{eq:growth}.}
\end{figure}

MI is possible in easy-plane ferromagnets, $\sigma=1$, only when
$|\bar{u}|>1$. In this case, wavelengths below the exchange length may
preclude the validity of the continuum approximation inherent in the
LL equation.  One potential resolution of such short-wave hydrodynamic
effects could lie in a discrete approach~\cite{Evans2014}. For
uniaxial ferromagnets, $\sigma=-1$, the speeds are always complex
except when $|\bar{n}|=1$ (vacuum) so that $\omega_\pm=\pm
|\mathbf{k}|^2$, i.e., the typical exchange spin wave dispersion
relation~\cite{Stancil2009}. In other words, ferromagnets with
perpendicular magnetic anisotropy are hydrodynamically unstable and
are analogous to attractive BECs~\cite{kevrekidis_defocusing_2015} and
focusing nonlinear optics~\cite{Kivshar2003} that support bright
solitons. In fact, magnetic solitons in ferromagnets with
perpendicular magnetic anisotropy have been predicted and
observed~\cite{Kosevich1990,Hoefer2010,Mohseni2013,Backes2015,Chung2018}. In
this case, the exponential growth rate of small perturbations is
$\Gamma(k)=\mathrm{Im}\{\omega_\pm(k)=\pm
k\sqrt{(1-\bar{n}^2)(\sigma-\bar{u}^2)+k^2}\}$, with
$k=|\mathbf{k}|$. Then, the maximum growth rate and associated
unstable wavenumber occur when $\Gamma'(k_\mathrm{max})=0$, given by
\begin{equation}
\label{eq:growth}
  \Gamma_\mathrm{max}=\frac{1}{2}k_c^2,\quad k_\mathrm{max}=\frac{1}{\sqrt{2}}k_c,\quad k_c=\sqrt{(1-\bar{n}^2)(1-\bar{u}^2)},
\end{equation}
where $k_c$ is the cutoff of the unstable band, see
Fig.~\ref{fig_extra}. Modulational instability is a long-wavelength
instability due to the interaction of nonlinear (PMA) and dispersive
(exchange) effects, which are well-known in many physical
environments~\cite{Whitham1974,zakharov_modulation_2009}.
%-------------------------------
%-------------------------------

%-------------------------------
\begin{figure*}[t]
\centering \includegraphics[trim={0.3in 2.6in 1in 0.3in}, clip, width=6in]{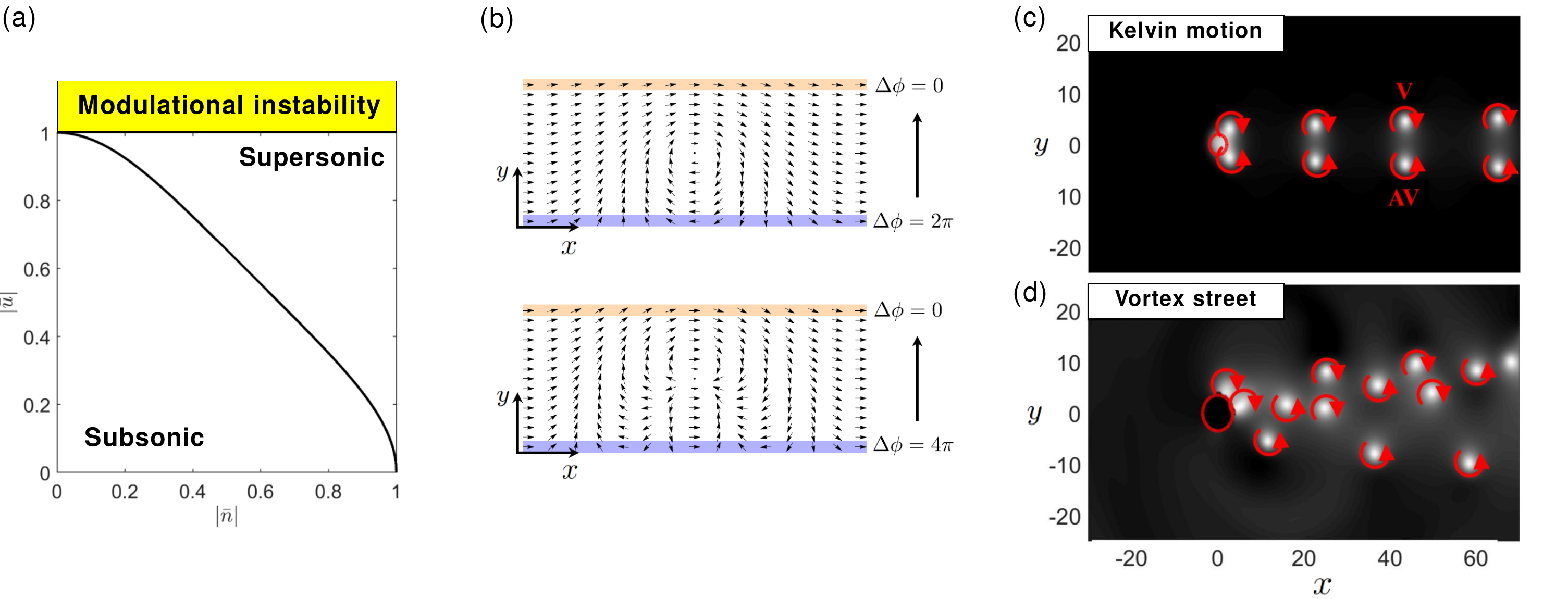}
\caption{ \label{fig2} (a) Density vs fluid velocity phase diagram for
  a UHS in an easy-plane ferromagnet. The solid black curve is the
  sonic curve where $\mathrm{M} = 1$. For $\bar{u}>1$, UHSs are
  modulationally unstable. (b) In-plane magnetization component with
  phase slips mediating a spatial change in the horizontal phase
  rotations, $\Delta\phi$, of a magnetic texture along the direction
  $y$. The top panel shows a transition from a $2\pi$ to $0$ rotation
  and the bottom panel a transition from a $4\pi$ to $0$
  rotation. Reprinted figure with permission from S. K. Kim and
  Y. Tserkovnyak, Phys. Rev. Lett. 116, 127201 (2016). Copyright
  (2016) by the American Physical Society. Examples of
  vortex-antivortex shedding from an impenetrable object exhibiting
  (c) Kelvin motion and (d) a von-K\'{a}rm\'{a}n-like vortex
  street~\cite{Iacocca2017b}}
\end{figure*}
%-------------------------------

\subsubsection{Modulationally stable regime}

For easy-plane ferromagnets, $\sigma=1$, the speeds of sound in Eq.~\eqref{eq:speed} are always real if $|\bar{u}|\leq1$. %The resulting Mach number is
In this case, a scenario similar to the Landau criterion is embodied by the subsonic to supersonic flow transition (when $s_-=0$ or $s_+=0$) quantified by a magnetic Mach number~\cite{Iacocca2017}
\begin{equation}
  \label{eq:Mach}
  \mathrm{M} =
  |\bar{u}|\sqrt{\frac{1+3\bar{n}^2}{1-\bar{n}^2}}.
\end{equation}

When $\mathrm{M}>1$, the flow is supersonic and UHSs can be destabilized by a persistent, large-amplitude disturbance, e.g., a physical obstacle~\cite{Iacocca2017b}. %and the UHS is unstable. A density vs fluid velocity phase diagram is shown in Fig.~\ref{fig2}(a).
When $\mathrm{M}<1$, the flow is laminar and the UHS stably flows around small obstacles. In the BEC and spin superfluid limits, Eq.~\eqref{eq:Mach} is equivalent to the Landau criterion for superfluidity~\cite{Sonin2017}. A density versus fluid velocity sub / supersonic phase diagram is shown in Fig.~\ref{fig2}(a). %However, if $\bar{u}>1$, i.e., the UHS wavelength approaches the exchange length, the speeds are complex and small fluctuation can exponentially grow, a scenario known as modulational instability (MI)~\cite{Whitham1974,zakharov_modulation_2009}. Physically, this case is at the limit of the applicability range of the LL equation and a discrete approach~\cite{Evans2014} should be used to appropriately resolve the dynamics.

%For uniaxial ferromagnets, $\sigma=+1$, the speeds are always complex and perturbations are unstable. This fact has led to the prediction and observation of stable solitons in these materials~\cite{Kosevich1990,Hoefer2010,Mohseni2013}.

%o	Vortex generation / phase slips
\subsection{Topological defects}

% usntable supersonic flow, how?
Because SDWs and UHSs are topologically protected, instabilities must necessarily lead to the nucleation of topological defects. Chirality can be unwound only through a phase singularity known generically as a phase-slip~\cite{Sonin2010,Kim2016,Kim2016b}, shown schematically in Fig.~\ref{fig2}(b). From a hydrodynamic perspective, such phase-slips are quantized vortices whose core is a density vacuum point, where $n=\pm1$~\cite{Iacocca2017b}. To conserve topology~\cite{papanicolaou_dynamics_1991}, singularities arise only in pairs whose aggregate topological number is zero, i.e., vortex-antivortex pairs with the same core polarity.

In the presence of finite-sized obstacles, vortex-antivortex pairs can appear even in subsonic conditions. This is because the deflected flow from a physical obstacle can locally become supersonic, resulting in vortex shedding, as is well known for classical fluids {undergoing a transition between laminar flow and the onset of turbulence}~\cite{Williamson1996,Leweke2016}. For easy-plane ferromagnets, such vortex-antivortex pair shedding has been numerically observed~\cite{Iacocca2017b} as well as their Kelvin motion~\cite{Papanicolaou1999} subject to a texture-induced Magnus force, shown in Fig.~\ref{fig2}(c). In addition, the simulations presented in Ref.~\cite{Iacocca2017b} showed evidence of superfluid-like behavior~\cite{Reeves2015,Frisch1992,Nore1993}, such as von-K\'{a}rm\'{a}n-like vortex streets~\cite{Sasaki2010,Kwon2016}, shown in Fig.~\ref{fig2}(d), and Mach cones with radiating wave-fronts ahead of the obstacle~\cite{Carusotto2006,Gladush2007}.

\section{Spin transport via spin hydrodynamics}

%One of the primary motivations to explore spin hydrodynamics in ferromagnets is to beat the exponential decay of spin-wave-mediated spin transport. In this section, we outline the general setting and discuss features of stabilized fluid-like, noncollinear magnetization states.

\subsection{Boundary value problem}

\begin{figure}[t]
\centering \includegraphics[trim={0.in 4.5in 2.5in 0in}, clip, width=2.5in]{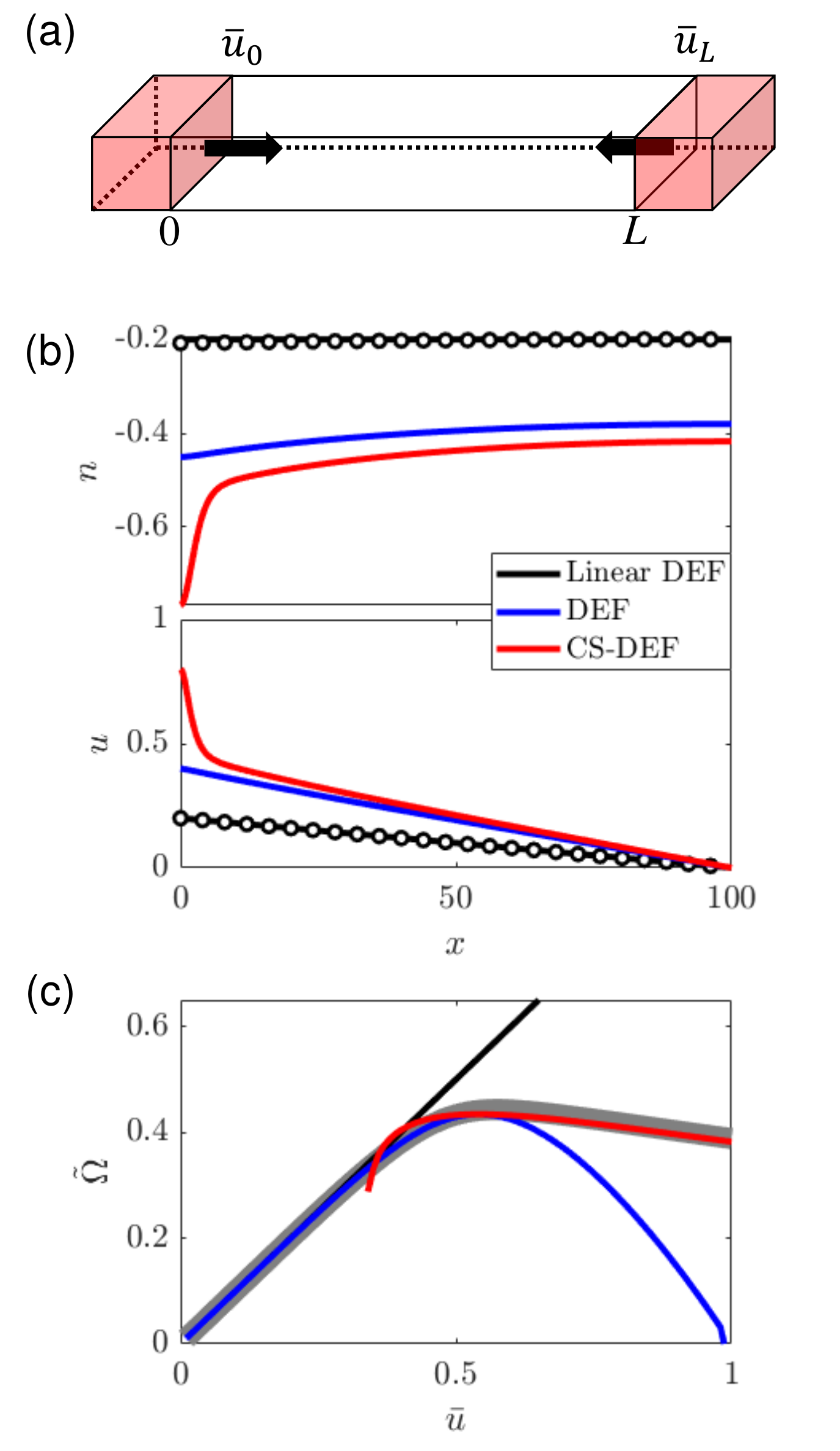}
\caption{ \label{fig3} (a) Schematic representation of a ferromagnetic channel subject to spin injection at both extrema. Spin injection can be realized in the red-shaded regions. (b) Density and fluid velocity profiles for the linear analytic solution (black curves) and numerical solution for a linear DEF (black circles), DEF (blue curves), and CS-DEF (red curves) in a ferromagnetic channel of length $L=100$ and $\alpha = 0.01$. (c) Analytically calculated frequency tunability for each solution. The full numerical solution is shown by a gray curve.}
\end{figure}

One of the primary motivations to explore spin hydrodynamics in ferromagnets is to beat the exponential decay of spin-wave-mediated spin transport {via noncollinear magnetization states}. Since typical free spin (open) boundary conditions $\partial\mathbf{m}/\partial\boldsymbol{\nu}=0$, where $\boldsymbol{\nu}$ is the boundary's outward normal, imply $u=0$, boundary control is needed to sustain noncollinear ($u\neq0$) spin hydrodynamic states. The simplest example of noncollinear magnetization states have been studied in easy-plane ferromagnetic channels subject to spin injection boundary conditions, shown schematically in Fig~\ref{fig3}(a). The boundaries represent magnetic / nonmagnetic interfaces that may be described as perfect spin sources or sinks~\cite{Iacocca2017d,Iacocca2019b} or imperfect boundaries, e.g., Pt/Py, subject to spin-mixing conductance~\cite{Takei2014} or spin pumping~\cite{Schneider2018}. Several physical methods for spin injection have been theoretically and experimentally proposed such as the spin-Hall effect~\cite{Stepanov2018,Takei2014}, spin-transfer torque~\cite{Iacocca2017d,Iacocca2019b,Schneider2018,Chen2014}, and the quantum spin-Hall effect~\cite{Yuan2018}.

%In the context of the DH formulation of Eqs.~\eqref{eq:ndot} and \eqref{eq:udot}, the boundary conditions take the form
%\begin{subequations}
%\label{eq:bc}
%\begin{eqnarray}
%  \label{eq:nbc}
%  \frac{\partial n(x=0)}{\partial \xi} = 0,&\quad&\frac{\partial n(x=L)}{\partial \xi} = 0,\\
%	\label{eq:ubc}
%	\mathbf{u}\cdot\hat{\boldsymbol{\xi}} = f_0(x),&\quad&\mathbf{u}\cdot\hat{\boldsymbol{\xi}} = f_L(x),
%\end{eqnarray}
%\end{subequations}
%where $f_0(x)$ and $f_L(x)$ represent the functions at the left edge, $x=0$ and right edge of a channel of dimensionless length $L$, $x=L$. For a ferromagnet, the dimensionless length has units of exchange length. 

Independent of the theoretical considerations for the boundary conditions, the equilibrium state is a solution to a boundary value problem (BVP). In other words, the solution necessarily spans the length of the channel and the profile of the solution is determined by the particular boundary conditions. This is in stark contrast to spin waves that are solutions to an initial value problem (IVP), i.e., where the magnet is relaxed into a particular state that is subject to either field or current perturbations and where the transverse physical boundaries play little role in the dominant spin wave characteristics~\cite{Chen2016b}. These fundamental differences between a BVP and an IVP are the reasons why noncollinear magnetization states can beat the exponential decay of spin waves.

For the sake of simplicity, we will consider in this review a perfect spin source coupled with open boundary conditions
\begin{subequations}
\label{eq:bc}
\begin{eqnarray}
  \label{eq:nbc}
  \frac{\partial n}{\partial x}(t,x=0) = 0,&\quad&\frac{\partial n}{\partial x}(t,x=L) = 0,\\
	\label{eq:ubc}
	u(t,x=0) = \bar{u}_0,&\quad&u(t,x=L) = \bar{u}_L,
\end{eqnarray}
\end{subequations}
where $\bar{u}_0$ and $\bar{u}_L$ represent fluid velocities at the extrema of a ferromagnetic channel of length $L$ in units of exchange length elongated in the $\hat{\mathbf{x}}$ direction so that the configuration has one-dimensional variation $\mathbf{u}(x)\cdot\hat{\mathbf{x}}=u(x)$. %This general setup is shown in Fig.~\ref{fig3}(a).
%-------------------------------
%-------------------------------

%o	Pumped nanowire / experiments: typical spin superfluids and also sims by brataas / josephson junction
%o	Contact solitons / screened superfluid / two-terminal
\subsection{Equilibrium solutions}

For the case of a channel subject to spin injection on one extremum, $\bar{u}_0\neq0$ and $\bar{u}_L=0$, both linear~\cite{Sonin2010,Takei2014} and nonlinear~\cite{Iacocca2017,Iacocca2019b} theory have been used to obtain solutions to the BVP that feature algebraic or near-algebraic spin transport. Furthermore, these equilibrium solutions consist of a coherent, single-frequency precessional state. %, similar to the solutions presented in Eq.~\eqref{eq:uhs_freq}.
Because magnetic damping necessarily must be included in the model, solutions to the BVP can be physically interpreted as dynamical modes that persist due to a nonlocal compensation of magnetic damping.

Linear, dispersionless theory based on Eqs.~\eqref{eq:nss} and \eqref{eq:uss} leads to a diffusion equation ultimately resulting in the so-called spin-superfluid~\cite{Sonin2010,Sonin2017,Sonin2019,Takei2014,Chen2014,Skarsvaag2015,Takei2015,Flebus2016}, alternatively linear dissipative exchange flow (DEF)~\cite{Iacocca2019b}
\begin{equation}
\label{eq:linearDEF}
  u(x) = \bar{u}_0\left(1-\frac{x}{L}\right),\quad\Omega=-n=\frac{\bar{u}_0}{\alpha L}.
\end{equation}

This solution is valid for sufficiently weak injection $|\bar{u}_0|\ll\mathrm{min}(1,\alpha L)$. The profile predicted by Eq.~\eqref{eq:linearDEF} is shown %An example solution for a ferromagnetic channel of length $L=100$ and $\alpha=0.01$ is shown 
in Fig.~\ref{fig3}(b) with solid black curves for a ferromagnetic channel of length $L=100$ and $\alpha=0.01$ subject to $\bar{u}_0=0.2$. The linear solution agrees with the numerically computed solution to the nonlinear BVP, shown by black circles. For $\bar{u}_0=0.4$, the density $n$ exhibits a nonlinear profile as shown by the numerical solution depicted by the solid blue curves. These states have been termed dissipative exchange flows (DEFs)~\cite{Iacocca2019b}. For even stronger injection, such that the conditions at the injection site are supersonic, a so-called contact soliton centered at $x=0$ is established, which then matches to a DEF solution. In this case, short-wave exchange dispersion is essential for the existence of a soliton. The numerical solution for this case is shown by the solid red curves when $\bar{u}_0=0.8$. The change in the spatial profiles of the solutions is accompanied by a nonlinear frequency tunability that was determined analytically for each case, as shown in Fig.~\ref{fig3}(c). The full numerical solution is shown by a gray curve.

Micromagnetic simulations have corroborated the existence of this solution both for ferromagnets~\cite{Iacocca2019b}, termed a contact-soliton DEF, and YIG~\cite{Schneider2018}, termed a soliton-screened spin superfluid. Detailed analytical derivations of all these modes under the DH formulation can be found in Ref.~\cite{Iacocca2019b}.

In the context of spin transport, the fluid velocity is proportional to the spin current, Eq.~\eqref{eq:Js}. It is important to recognize that the net spin transport is set by the boundaries. In the ideal case considered here, the net spin transport is zero because there is no spin sink at the right edge. By taking boundaries into account, spin transport is possible~\cite{Takei2014}. The extreme example is that of a channel where both extrema are subject to spin injection, $\bar{u}_0\neq0$ and $\bar{u}_L\neq0$. In this case, it has been shown that static solutions similar to SDWs can be stabilized in the context of linear theory~\cite{Hill2018}, when $\bar{u}_0\sim \bar{u}_L$ so that $n=0$. An analogy to spin Josephson junctions was drawn in Ref.~\cite{Hill2018}, owing to the spatially homogeneous fluid velocity or spin current supported in these states.

Finally, it is worth noting that spin current transport mediated by the precessional frequency of the equilibrium solution is also possible. In this case, the precessional frequency can give rise to a spin current in an adjacent spin reservoir via spin pumping~\cite{Tserkovnyak2002b,Brataas2012}.% In fact, the precessional frequency is the only spatially homogeneous parameter in the equilibrium solutions~\cite{Iacocca2019b}.

%o	Symmetry-breaking effects
\subsection{Symmetry-breaking effects}

When the easy-plane symmetry of the system is broken, additional terms that are functions of $\Phi$ must be included in Eqs.~\eqref{eq:ndot} and \eqref{eq:udot}~\cite{Iacocca2017d}. This can occur, for example, due to magnetocrystalline anisotropy in soft magnets, nonlocal dipole fields in the LL equation~\cite{Sprenger2019}, or in-plane applied fields.

For the case of magnetocrystalline anisotropy, the resulting equations of motion take the form of a sine-Gordon equation in the {weak-injection} regime~\cite{Sonin2010,Iacocca2017d}. Physically, this implies that equilibrium solutions are trains of N\'{e}el domain walls with the same chirality. As a consequence, a threshold current exists, proportional to the energy necessary to tilt the magnetization towards the hard axis~\cite{Sonin2010,Iacocca2017d}. The threshold is linearly proportional to the anisotropy field and the channel length for channels shorter than the domain wall width and saturates at a value proportional to the square root of the anisotropy field for long channels~\cite{Iacocca2017d}. The threshold also induces a reduction in the precessional frequency, reducing the efficiency of spin pumping into a spin reservoir. The predicted threshold spin current density for typical Permalloy material parameters is on the order of $10^{-4}$~J/m$^2$. These values can be obtained by rather large charge current densities on the order of $10^{12}$~A/m$^2$ via spin Hall or spin torque effects. While such charge current densities in principle achievable, new materials with large spin Hall angle~\cite{Keller2019} may play an important role to further decrease this figure-of-merit.

Symmetry breaking due to nonlocal dipole fields has been argued to preclude the stabilization of linear DEFs or spin superfluids beyond the distance of an exchange length~\cite{Skarsvaag2015}. However, micromagnetic simulations~\cite{Iacocca2017,Iacocca2017d,Iacocca2019b,Schneider2018} have demonstrated that SDWs and linear DEF solutions can be stabilized even in the presence of nonlocal dipole fields. Nonlocal dipole has also been studied in the context of %spatially homogeneous states and its perturbative effect on SDWs where a band diagram, including band gaps, is induced~\cite{Sprenger2019}. 
the spatial modulation of SDWs~\cite{Sprenger2019}. Such a modulation induces a spin wave band diagram, including band gaps, that opens opportunities for noncollinear magnetization states as reprogrammable magnonic crystals~\cite{Krawczyk2014}.% However, micromagnetic simulations including nonlocal dipole fields~\cite{Iacocca2017,Iacocca2017d,Iacocca2019b,Schneider2018} have demonstrated qualitatively similar results to those predicted analytically in the symmetric case.

Finally, we remark that finite temperature micromagnetic simulations have also demonstrated the robustness of UHSs and SDWs to stochastic thermal fields~\cite{Iacocca2017}.

%o Defects carried by spin superfluids
\subsection{Other applications}

Theoretical studies of current-induced states in ferromagnetic channels have also explored their interactions with other quasi-particles. A temperature gradient along the ferromagnetic channel can excite spin waves~\cite{Uchida2008} that interact with noncollinear magnetization states in a manner described by a two-fluid model composed of a superfluid and normal fluid~\cite{Flebus2016,Tserkovnyak2016}. %The thickness of realistic ferromagnetic nanowires gives rise to nonlocal dipole fields that modulate current-induced SDWs~\cite{Sprenger2019}. Such a modulation produces a spin wave band diagram, opening opportunities for noncollinear magnetization states as reprogrammable magnonic crystals~\cite{Krawczyk2014}.
It was also shown that thermally activated vortices on a noncollinear magnetization state~\cite{Kim2016b} diffuse along the ferromagnetic channel and can induce a spin current in an adjacent spin reservoir~\cite{Kim2015}.

\section{Perspectives and challenges}

%o	Some brief concluding remarks.
The field of spin hydrodynamics is undergoing rapid theoretical
development that has lead to realistic predictions in various
contexts. While this review has emphasized ferromagnetic materials,
spin hydrodynamics have been extended beyond, including spin transport
in antiferromagnets~\cite{Takei2014b,Qaiumzadeh2017}, noncollinear
antiferromagnets~\cite{Yamane2019}, and amorphous
materials~\cite{Ochoa2018}.  These early works have already
demonstrated that the hydrodynamic perspective of nonlinear dynamics
in magnetic materials results from a distinctly new interpretation
that yields significant predictive value.  This points toward growing
interest in this field of research and a number of intriguing research
directions, some of which have already been touched upon in this
review.  Examples include the exploration of sub-exchange length
hydrodynamics, three-dimensional effects, and the dispersive
hydrodynamics of more general magnetic order.

Despite the many theoretical predictions, a conclusive experimental observation of spin hydrodynamics in ferromagnets remains elusive. However, recent experimental evidence in antiferromagnets~\cite{Stepanov2018,Yuan2018} is promising for the development of experimental spin hydrodynamics in ferromagnets.

Perspectives for spin hydrodynamics may be found by cross-fertilizing
with the broader field of fluid dynamics. For example, a hydrodynamic
analysis of the spatially varying magnetization of ferrimagnetic
GdFeCo subject to ultrafast optical pumping suggests the appearance of
localized, topological textures at picosecond timescales as well as
the onset of turbulent spin transport~\cite{Iacocca2019}. More
fundamentally, the role of magnetic damping and its fluid
interpretation may lead to additional insights on losses in magnetic
texture dynamics.  Fruitful future research directions also include
time-dependent and emergent phenomena~\cite{Buzzi2018} such as rare
events e.g., rogue waves~\cite{Chabchoub2016}, and the appearance of
nonlinear structures known as dispersive shock waves~\cite{El2016} in
both hydrodynamically (modulationally) stable and unstable magnetic
configurations.  We envision a bright future for the rapidly growing
field of spin hydrodynamics.

\section{Acknowledgments}

E.I. and M.A.H. acknowledge support from the U.S. Department of Energy, Office of Science, Office of Basic Energy Sciences under Award Number DE-SC0017643. M.A.H. partially supported by NSF CAREER DMS-1255422.

\bibliographystyle{elsarticle-num}

\begin{thebibliography}{10}
\expandafter\ifx\csname url\endcsname\relax
  \def\url#1{\texttt{#1}}\fi
\expandafter\ifx\csname urlprefix\endcsname\relax\def\urlprefix{URL }\fi
\expandafter\ifx\csname href\endcsname\relax
  \def\href#1#2{#2} \def\path#1{#1}\fi

\bibitem{Roadmap2017}
D.~Sander, S.~O. Valenzuela, D.~Makarov, C.~H. Marrows, E.~E. Fullerton,
  P.~Fischer, J.~McCord, P.~Vavassori, S.~Mangin, P.~Pirro, B.~Hillebrands,
  A.~D. Kent, T.~Jungwirth, O.~Gutfleisch, C.~G. Kim, A.~Berger, The 2017
  magnetism roadmap, Journal of Physics D: Applied Physics 50~(36) (2017)
  363001.

\bibitem{Hoffmann2007}
A.~Hoffmann, Pure spin-currents, Phys. Stad. Sol. 11 (2007) 4236.

\bibitem{Jungwirth2012}
T.~Jungwirth, J.~Wunderlich, K.~Olejn\'{i}k, Spin hall effect devices, Nature
  Materials 11 (2012) 382.

\bibitem{Manchon2015}
A.~Manchon, H.~C. Koo, J.~Nitta, S.~M. Frolov, R.~A. Duine, New perspectives
  for rashba spin-orbit coupling, Nature Materials 14 (2015) 871.

\bibitem{Hoffmann2013}
A.~Hoffmann, Spin hall effects in metals, IEEE Advances on magnetics 49~(10)
  (2013) 5172.

\bibitem{Humphries2017}
A.~M. Humphries, T.~Wang, E.~R.~J. Edwards, S.~R. Allen, J.~M. Shaw, H.~T.
  Nembach, J.~Q. Xiao, T.~J. Silva, X.~Fan, Observation of spin-orbit effects
  with spin rotation symmetry, Nature Communications 8 (2017) 911.

\bibitem{Kimata2019}
M.~Kimata, H.~Chen, K.~Kondory, S.~Sugimoto, P.~K. Muduli, M.~Ikhlas, Y.~Omori,
  T.~Tomita, A.~H. MacDonald, S.~Nakatsuji, T.~Otani, Magnetic and magnetic
  inverser spin hall effects in a non-collinear antiferromagnet, Nature 565
  (2019) 627.

\bibitem{White2007}
R.~White, Quantum theory of magnetism, Springer, 2007.

\bibitem{Slonczewski1996}
J.~C. Slonczewski, Current-driven excitation of magnetic multilayers, J. Magn.
  Magn. Mater. 159~(1-2) (1996) L1 -- L7.

\bibitem{berger1996}
L.~Berger, Emission of spin waves by a magnetic multilayer traversed by a
  current, Phys. Rev. B 54~(13) (1996) 9353--9358.

\bibitem{Tserkovnyak2002b}
Y.~Tserkovnyak, A.~Brataas, G.~E.~W. Bauer, Enhanced gilbert damping in thin
  ferromagnetic films, Phys. Rev. Lett. 88 (2002) 117601.

\bibitem{Brataas2012}
A.~Brataas, A.~D. Kent, H.~Ohno, Current-induced torques in magnetic materials,
  Nature Materials 11 (2012) 372.

\bibitem{Suhl1998}
H.~Suhl, Theory of the magnetic damping constant, Magnetics, IEEE Transactions
  on 34~(4) (1998) 1834 --1838.

\bibitem{Madami2011}
M.~Madami, S.~Bonetti, G.~Consolo, S.~Tacchi, G.~Carlotti, G.~Gubbiotti, F.~B.
  Mancoff, M.~A. Yar, J.~\AA{}kerman, Direct observation of a propagating spin
  wave induced by spin-transfer torque, Nature Nanotechnology 6 (2011)
  635--638.

\bibitem{Gilbert2004}
T.~L. Gilbert, A phenomenological theory of damping in ferromagnetic materials,
  Magnetics, IEEE Transactions on 40~(6) (2004) 3443 -- 3449.

\bibitem{Chumak2015}
A.~V. Chumak, V.~I. Vasyuchka, A.~A. Serga, B.~Hillebrands, Magnon spintronics,
  Nature physics 11 (2015) 453.

\bibitem{Cornelissen2015}
L.~J. Cornelissen, J.~Liu, R.~A. Duine, J.~B. Youssef, B.~J. van Wees,
  Long-distance transport of magnon spin information in a magnetic insulator at
  room temperature, Nature Physics 11 (2015) 1022.

\bibitem{Liu2018}
C.~Liu, J.~Chen, T.~Liu, F.~Heimbach, H.~Yu, Y.~Xiao, J.~Hu, M.~Liu, H.~Chang,
  T.~Stueckler, S.~tu, Y.~Zhang, Y.~Zhang, P.~Gao, Z.~Liao, D.~Yu, K.~Xia,
  N.~Lei, W.~Zhao, M.~Wu, Long-distance propagation of short-wavelength spin
  waves, Nature Communications 9 (2018) 738.

\bibitem{Wesenberg2017}
D.~Wesenberg, T.~Liu, D.~Balzar, M.~Wu, B.~L. Zink, Long-distance spin
  transport in a disordered magnetic insulator, Nature Physics 13 (2017) 987.

\bibitem{Lebrun2018}
R.~Lebrun, A.~Ross, S.~A. Bender, A.~Qaiumzadeh, L.~Baldrati, J.~Cramer,
  A.~Brataas, R.~A. Duine, M.~Kl\"{a}ui, Tunable long-distance spin transport
  in crystalline antiferromagnetic iron oxide, Nature Physics 561 (2018) 222.

\bibitem{Stepanov2018}
P.~Stepanov, S.~Che, D.~Shcherbakov, J.~Yang, R.~Chen, K.~Thilahar, G.~Voigt,
  M.~W. Bockrath, D.~Smirnov, K.~Watanabe, T.~Taniguchi, R.~K. Lake, Y.~Barlas,
  A.~H. MacDonald, C.~N. Lau, Long-distance spin transport through a graphene
  quantum hall antiferromagnet, Nature Physics 14 (2018) 907.

\bibitem{Yuan2018}
W.~Yuan, Q.~Zhu, T.~Su, Y.~Yao, W.~Xing, Y.~Chen, Y.~Ma, X.~Lin, J.~Shi,
  R.~Shindou, X.~C. Xie, W.~Han, Experimental signatures of spin superfluid
  ground state in canted antiferromagnet cr$_2$o$_3$ via nonlocal spin
  transport, Science Advances 4 (2018) eaat1098.

\bibitem{Landau1953}
L.~D. Landau, E.~Lifshitz, On the theory of the dispersion of magnetic
  permeability in ferromagnetic bodies, Phys. Z. Sowjet. 8 (1953) 153.

\bibitem{Bruno1989}
P.~Bruno, J.~P. Renard, Magnetic surface anisotropy of transition metal
  ultrathin films, Applied Physics A: Materials Science \&amp; Processing 49
  (1989) 499--506, 10.1007/BF00617016.

\bibitem{Iacocca2017}
E.~Iacocca, T.~J. Silva, M.~A. Hoefer, Breaking of galilean invariance in the
  hydrodynamic formulation of ferromagnetic thin films, Phys. Rev. Lett. 118
  (2017) 017203.

\bibitem{Halperin1969}
B.~Halperin, P.~Hohenberg, Hydrodynamic theory of spin waves, Physical Review
  188~(2) (1969) 898--918.

\bibitem{papanicolaou_dynamics_1991}
N.~Papanicolaou, T.~Tomaras, Dynamics of magnetic vortices, Nuclear Physics B
  360~(2-3) (1991) 425--462.

\bibitem{Sonin2010}
E.~B. Sonin, Spin currents and spin superfluidity, Advances in Physics 59~(3)
  (2010) 181 -- 255.

\bibitem{Konig2001}
J.~K\"onig, M.~C. B\o{}nsager, A.~H. MacDonald, Dissipationless spin transport
  in thin film ferromagnets, Phys. Rev. Lett. 87 (2001) 187202.

\bibitem{Gruner1994}
G.~Gr\"uner, The dynamics of spin-density waves, Rev. Mod. Phys. 66 (1994)
  1--24.

\bibitem{Iacocca2017b}
E.~Iacocca, M.~A. Hoefer, Vortex-antivortex proliferation from an obstacle in
  thin film ferromagnets, Phys. Rev. B 95 (2017) 134409.

\bibitem{Landau1987}
L.~D. Landau, E.~M. Lifshitz, Fluid Mechanics, Pergamon press, 1987.

\bibitem{Bruno2005}
P.~Bruno, V.~K. Dugaev, Equilibrium spin currents and the magnetoelectric
  effect in magnetic nanostructures, Phys. Rev. B 72 (2005) 241302.

\bibitem{Iacocca2019}
E.~Iacocca, T.-M. Liu, A.~H. Reid, Z.~Fu, S.~Ruta, P.~W. Granitzka, E.~Jal,
  S.~Bonetti, A.~X. Gray, C.~E. Graves, R.~Kukreja, Z.~Chen, D.~J. Higley,
  T.~Chase, L.~Le~Guyader, K.~Hirsch, H.~Ohldag, W.~F. Schlotter, G.~L.
  Dakovski, G.~Coslovich, M.~C. Hoffmann, S.~Carron, A.~Tsukamoto, A.~Kirilyuk,
  A.~V. Kimel, T.~Rasing, J.~St\"{o}hr, R.~F.~L. Evans, T.~Ostler, R.~W.
  Chantrell, M.~A. Hoefer, T.~J. Silva, , H.~A. D\"{u}rr, Spin-current-mediated
  rapid magnon localization and coalescence after ultrafast optical pumping of
  ferrimagnetic alloys, Nature Communications 10 (2019) 1756.

\bibitem{Stancil2009}
D.~Stancil, A.~Prabhakar, Spin waves: Theory and applications, Springer, 2009.

\bibitem{Pethick2002}
C.~Pethick, H.~Smith, Bose-Einstein condensation in dilute gasses, Cambridge
  University Press, 2002.

\bibitem{Qu2016}
C.~Qu, L.~P. Pitaevskii, S.~Stringari, Magnetic solitons in a binary
  bose-einstein condensate, Phys. Rev. Lett. 116 (2016) 160402.

\bibitem{Congy2016}
T.~Congy, A.~M. Kamchatnov, N.~Pavloff, {Dispersive hydrodynamics of nonlinear
  polarization waves in two-component Bose-Einstein condensates}, SciPost Phys.
  1 (2016) 006.

\bibitem{El2016}
G.~El, M.~Hoefer, Dispersive shock waves and modulation theory, Physica D, to
  appear.

\bibitem{Ivanov2017}
S.~K. Ivanov, A.~M. Kamchatnov, T.~Congy, N.~Pavloff, Solution of the riemann
  problem for polarization waves in a two-component bose-einstein condensate,
  Phys. Rev. E 96 (2017) 062202.

\bibitem{Ivanov2018}
S.~K. Ivanov, A.~M. Kamchatnov, Simple waves in a two-component bose-einstein
  condensate, Phys. Rev. E 97 (2018) 042208.

\bibitem{Landau1941}
L.~D. Landau, Theory of the superfluidity of helium {II}, J. Phys. USSR 5
  (1941) 71.

\bibitem{Sonin2017}
E.~B. Sonin, Spin superfluidity and spin waves in yig films, Phys. Rev. B 95
  (2017) 144432.

\bibitem{Sonin2019}
E.~B. Sonin, Superfluid spin transport in ferro- and antiferromagnets, Phys.
  Rev. B 99 (2019) 104423.

\bibitem{Whitham1974}
G.~B. Whitham, Linear and nonlinear waves, John Wiley \& Sons Inc, 1974.

\bibitem{zakharov_modulation_2009}
V.~Zakharov, L.~Ostrovsky, Modulation instability: {The} beginning, Physica D
  238 (2009) 540--548.

\bibitem{Evans2014}
R.~F.~L. Evans, W.~J. Fan, P.~Chureemart, T.~A. Ostler, M.~A.~A. Allis, R.~W.
  Chantrell, Atomistic spin model simulations of magnetic nanomaterials, J.
  Phys.: Condens. Matter 26 (2014) 103202.

\bibitem{kevrekidis_defocusing_2015}
P.~G. Kevrekidis, D.~J. Frantzeskakis, R.~Carretero-Gonz\'{a}lez, The
  {Defocusing} {Nonlinear} {Schr\"{o}dinger} {Equation}, SIAM, Philadelphia,
  2015.

\bibitem{Kivshar2003}
Y.~Kivshar, G.~Agrawal, Optical solitons, Elsevier, 2003.

\bibitem{Kosevich1990}
A.~Kosevich, B.~Ivanov, A.~Kovalev, Magnetic solitons, Physics Reports
  194~(3–4) (1990) 117 -- 238.

\bibitem{Hoefer2010}
M.~A. Hoefer, T.~J. Silva, M.~W. Keller, Theory for a dissipative droplet
  soliton excited by a spin torque nanocontact, Phys. Rev. B 82 (2010) 054432.

\bibitem{Mohseni2013}
S.~M. Mohseni, S.~R. Sani, J.~Persson, T.~N.~A. Nguyen, S.~Chung,
  Y.~Pogoryelov, P.~K. Muduli, E.~Iacocca, A.~Eklund, R.~K. Dumas, S.~Bonetti,
  A.~Deac, M.~A. Hoefer, J.~\AA{}kerman, Spin torque-generated
  magnetic droplet solitons, Science 339~(6125) (2013) 1295--1298.

\bibitem{Backes2015}
D.~Backes, F.~Maci\`a, S.~Bonetti, R.~Kukreja, H.~Ohldag, A.~D. Kent, Direct
  observation of a localized magnetic soliton in a spin-transfer nanocontact,
  Phys. Rev. Lett. 115 (2015) 127205.

\bibitem{Chung2018}
S.~Chung, Q.~T. Le, M.~Ahlberg, A.~A. Awad, M.~Weigand, I.~Bykova, R.~Khymyn,
  M.~Dvornik, H.~Mazraati, A.~Houshang, S.~Jiang, T.~N.~A. Nguyen, E.~Goering,
  G.~Sch\"utz, J.~Gr\"afe, J.~\AA{}kerman, Direct observation of zhang-li
  torque expansion of magnetic droplet solitons, Phys. Rev. Lett. 120 (2018)
  217204.

\bibitem{Kim2016}
S.~K. Kim, Y.~Tserkovnyak, Topological effects on quantum phase slips in
  superfluid spin transport, Phys. Rev. Lett. 116 (2016) 127201.

\bibitem{Kim2016b}
S.~K. Kim, S.~Takei, Y.~Tserkovnyak, Thermally activated phase slips in
  superfluid spin transport in magnetic wires, Phys. Rev. B 93 (2016) 020402.

\bibitem{Williamson1996}
C.~H.~K. Williamson, Vortex {Dynamics} in the {Cylinder} {Wake}, Annual Review
  of Fluid Mechanics 28~(1) (1996) 477--539.

\bibitem{Leweke2016}
T.~Leweke, S.~Le~Diz\`{e}s, C.~H.~K. Williamson, Dynamics and instabilityies of
  vortex pairs, Annual reviews of Fluid Mechanics 48 (2016) 506--541.

\bibitem{Papanicolaou1999}
N.~Papanicolaou, P.~N. Spathis, Semitopological solitons in planar
  ferromagnets, Nonlinearity 12 (1999) 285.

\bibitem{Reeves2015}
M.~T. Reeves, T.~P. Billam, B.~P. Anderson, A.~S. Bradley, Identifying a
  superfluid reynolds number via dynamical similarity, Phys. Rev. Lett. 114
  (2015) 155302.

\bibitem{Frisch1992}
T.~Frisch, Y.~Pomeau, S.~Rica, Transition to dissipation in a model of
  superflow, Phys. Rev. Lett. 69 (1992) 1644--1647.

\bibitem{Nore1993}
C.~Nore, M.~Brachet, S.~Fauve, Numerical study of hydrodynamics using the
  nonlinear schrÃ¶dinger equation, Physica D: Nonlinear Phenomena 65~(1) (1993)
  154 -- 162.

\bibitem{Sasaki2010}
K.~Sasaki, N.~Suzuki, H.~Saito, B\'enard\char21{}von k\'arm\'an vortex street
  in a bose-einstein condensate, Phys. Rev. Lett. 104 (2010) 150404.

\bibitem{Kwon2016}
W.~J. Kwon, J.~H. Kim, S.~W. Seo, Y.~Shin, Observation of von k\'arm\'an vortex
  street in an atomic superfluid gas, Phys. Rev. Lett. 117 (2016) 245301.

\bibitem{Carusotto2006}
I.~Carusotto, S.~X. Hu, L.~A. Collins, A.~Smerzi, Bogoliubov-\ifmmode
  \check{C}\else \v{C}\fi{}erenkov radiation in a bose-einstein condensate
  flowing against an obstacle, Phys. Rev. Lett. 97 (2006) 260403.

\bibitem{Gladush2007}
Y.~G. Gladush, G.~A. El, A.~Gammal, A.~M. Kamchatnov, Radiation of linear waves
  in the stationary flow of a bose-einstein condensate past an obstacle, Phys.
  Rev. A 75 (2007) 033619.

\bibitem{Iacocca2017d}
E.~Iacocca, T.~J. Silva, M.~A. Hoefer, Symmetry-broken dissipative exchange
  flows in thin-film ferromagnets with in-plane anisotropy, Phys. Rev. B 96
  (2017) 134434.

\bibitem{Iacocca2019b}
E.~Iacocca, M.~A. Hoefer, Hydrodynamic description of long-distance spin
  transport through noncollinear magnetization states: Role of dispersion,
  nonlinearity, and damping, Phys. Rev. B 99 (2019) 184402.

\bibitem{Takei2014}
S.~Takei, Y.~Tserkovnyak, Superfluid spin transport through easy-plane
  ferromagnetic insulators, Phys. Rev. Lett. 112 (2014) 227201.

\bibitem{Schneider2018}
T.~Schneider, D.~Hill, A.~K\'{a}kay, K.~Lenz, J.~Lindner, J.~Fassbender,
  P.~Upadhyaya, Y.~Liu, K.~Wang, Y.~Tserkovnyak, I.~N. Krivorotov, I.~Barsukov,
  Self-stabilizing spin superfluid, arXiv:1811.09369.

\bibitem{Chen2014}
H.~Chen, A.~D. Kent, A.~H. MacDonald, I.~Sodemann, Nonlocal transport mediated
  by spin supercurrents, Phys. Rev. B 90 (2014) 220401.

\bibitem{Chen2016b}
T.~Chen, R.~K. Dumas, A.~Eklund, P.~K. Muduli, A.~Houshang, A.~A. Awad,
  P.~D\"{u}rrenfeld, B.~G. Malm, A.~Rusu, J.~\AA{}kerman, Spin-torque and
  spin-hall nano-oscillators, Proceedings of the IEEE 104~(10) (2016)
  1919--1945.

\bibitem{Skarsvaag2015}
H.~Skarsv\aa{}g, C.~Holmqvist, A.~Brataas, Spin superfluidity and long-range
  transport in thin-film ferromagnets, Phys. Rev. Lett. 115 (2015) 237201.

\bibitem{Takei2015}
S.~Takei, Y.~Tserkovnyak, Nonlocal magnetoresistance mediated by spin
  superfluidity, Phys. Rev. Lett. 115 (2015) 156604.

\bibitem{Flebus2016}
B.~Flebus, S.~A. Bender, Y.~Tserkovnyak, R.~A. Duine, Two-fluid theory for spin
  superfluidity in magnetic insulators, Phys. Rev. Lett. 116 (2016) 117201.

\bibitem{Hill2018}
D.~Hill, S.~K. Kim, Y.~Tserkovnyak, Spin-torque-biased magnetic strip:
  Nonequilibrium phase diagram and relation to long josephson junctions, Phys.
  Rev. Lett. 121 (2018) 037202.

\bibitem{Sprenger2019}
P.~{Sprenger}, M.~A. {Hoefer}, E.~{Iacocca}, Magnonic band structure
  established by chiral spin-density waves in thin-film ferromagnets, IEEE
  Magnetics Letters 10 (2019) 4501605.

\bibitem{Keller2019}
M.~W. Keller, K.~S. Gerace, M.~Arora, E.~K. Delczeg-Czirjak, J.~M. Shaw, T.~J.
  Silva, Near-unity spin hall ratio in
  $\mathrm{N}{\mathrm{i}}_{x}\mathrm{C}{\mathrm{u}}_{1\ensuremath{-}x}$ alloys,
  Phys. Rev. B 99 (2019) 214411.

\bibitem{Krawczyk2014}
M.~Krawczyk, D.~Grundler, Review and prospects of magnonic crystals and devices
  with reprogrammable band structure, Journal of Physics: Condensed Matter
  26~(12) (2014) 123202.

\bibitem{Uchida2008}
K.~Uchida, S.~Takahashi, K.~Harii, J.~Ieda, W.~Koshibae, K.~Ando, S.~Maekawa,
  E.~Saitoh, Observation of the spin seebeck effect, Nature 455 (2008) 778.

\bibitem{Tserkovnyak2016}
Y.~Tserkovnyak, S.~A. Bender, R.~A. Duine, B.~Flebus, Bose-einstein
  condensation of magnons pumped by the bulk spin seebeck effect, Phys. Rev. B
  93 (2016) 100402.

\bibitem{Kim2015}
S.~K. Kim, S.~Takei, Y.~Tserkovnyak, Topological spin transport by brownian
  diffusion of domain walls, Phys. Rev. B 92 (2015) 220409.

\bibitem{Takei2014b}
S.~Takei, B.~I. Halperin, A.~Yacoby, Y.~Tserkovnyak, Superfluid spin transport
  through antiferromagnetic insulators, Phys. Rev. B 90 (2014) 094408.

\bibitem{Qaiumzadeh2017}
A.~Qaiumzadeh, H.~Skarsv\aa{}g, C.~Holmqvist, A.~Brataas, Spin superfluidity in
  biaxial antiferromagnetic insulators, Phys. Rev. Lett. 118 (2017) 137201.

\bibitem{Yamane2019}
J.~S. Y.~Yamane, O.~Gomonay, Dyanmics of noncollinear antiferromagnetic
  textures driven by spin current injection, arXiv:1901.05684.

\bibitem{Ochoa2018}
H.~Ochoa, R.~Zarzuela, Y.~Tserkovnyak, Spin hydrodynamics in amorphous magnets,
  Phys. Rev. B 98 (2018) 054424.

\bibitem{Buzzi2018}
M.~Buzzi, M.~F\"{o}rst, R.~Mankowsky, A.~Cavalleri, Probing dynamics in quantum
  materials with femtosecond x-rays, Nature Reviews Materials 3 (2018) 299.

\bibitem{Chabchoub2016}
A.~Chabchoub, M.~Onorato, N.~Akhmediev, Hydrodynamic envelope solitons and
  breathers, Springer, 2016, Ch.~3, pp. 55--87.

\end{thebibliography}

\end{document}